\documentclass[usenatbib,useAMS]{mn2e}
\usepackage{epsfig,mncite}
\usepackage[varg]{txfonts}

\def\msun{{\rm M_{\odot}}}

\title [Planetesimal Formation ]
{Limits on the location of planetesimal formation in  self-gravitating protostellar discs}
\author[C.J.~Clarke and G. Lodato]{C.J.~Clarke$^1$ and G. Lodato$^2$\\
$^1$Institute of Astronomy, Madingley Rd, Cambridge, CB3 0HA, UK\\
$^{2}$Dipartimento di Fisica, Universit\`a degli studi di Milano, Via Celoria 16, Milano, I-20133, Italy}
 
\date{Submitted:}
\pagerange{\pageref{firstpage}--\pageref{lastpage}} \pubyear{2009}

\begin{document}
\def\lta{\mathrel{\spose{\lower 3pt\hbox{$\mathchar"218$}}
     \raise 2.0pt\hbox{$\mathchar"13C$}}}
\def\gta{\mathrel{\spose{\lower 3pt\hbox{$\mathchar"218$}}
     \raise 2.0pt\hbox{$\mathchar"13E$}}}
\def\Msun{{\rm M}_\odot}
\def\msun{{\rm M}_\odot}
\def\Rsun{{\rm R}_\odot}
\def\Lsun{{\rm L}_\odot}
\def\19{GRS~1915+105}
\label{firstpage}
\maketitle

\begin{abstract}

In this Letter we show that if planetesimals form in spiral features
in self-gravitating discs, as previously suggested by the idealised
simulations of Rice et al, then in
realistic protostellar discs, this process will be restricted to the outer
regions of the disc (i.e. at radii in excess of several tens of A.U.).
This restriction relates to the requirement that dust has to be concentrated
in spiral features on a timescale that is less than the (roughly dynamical)
lifetime of such features, and that such rapid accumulation requires
spiral features whose fractional amplitude is not much less than unity.
This in turn requires that the  cooling timescale of the gas is relatively 
short, which restricts the process to the outer disc. We point out that
the efficient conversion of a large fraction of the primordial dust
in the disc into planetesimals could rescue this material from the well
known problem of rapid inward migration at a $\sim$ metre size scale 
and that in principle the collisional evolution of these objects could help
to re-supply small dust to the protostellar disc. We also point out the
possible implications of this scenario for the location of planetesimal
belts inferred in debris discs around main sequence stars, but stress that 
further dynamical studies are required in order to establish whether the
disc retains a memory of the initial site of planetesimal creation.

\end{abstract}

\begin{keywords}
accretion, accretion discs - planetary systems: formation - hydrodynamics - instabilities
\end{keywords}

\section{Introduction}

A much debated aspect of protostellar disc evolution concerns the way that
the solid component of the disc (for which the range of grain sizes is initially
in the sub-micron range as in the interstellar medium; Dullemond et al., 2007) 
is assembled into larger bodies. This process is not only evidenced by the 
range of rocky/icy bodies (terrestrial planets, comets, asteroids) in our Solar System, 
but also by the existence of dusty debris around young main sequence stars, 
which are interpreted in terms of the grinding down of  a reservoir of rocky/icy 
planetesimals of km size or more (Wyatt \& Dent, 2002, Greaves et al., 2004, Wyatt, 2008). 
Clearly, therefore,  large rocky/icy bodies are in place by an age of around 10 
Myr (corresponding to the  youngest debris discs) but this does not in itself 
provide any constraints about {\it when} these objects are assembled during the 
preceding, gas rich phase of disc evolution. Spectral evidence of grain growth 
has however been accumulating in recent years, as the form of the spectral energy
distribution at mm wavelengths and beyond appears to require the existence 
of emitting particles of at least cm scales in some objects (e.g., Testi et al. 2003). 
The recent detection of emission at a wavelength of 10 cm in the protostellar disc 
system HL Tau is particularly striking in this regard (Greaves et al., 2008): although 
the extent to which this emission is contaminated by non-thermal (free-free) emission 
is debatable, the detection of thermal emission at this wavelength requires 
substantial grain growth (to $> 10$ cm) in a system that is still young 
($<$ a Myr old) and where the disc is massive ($> 0.1 M_\odot$). Apparently, then, 
grain growth is already under way in early evolutionary stages when the disc 
is still strongly {\it self-}gravitating.

Grain growth however presents a potential problem for the retention of solid material 
in the disc, since objects of metre size  are subject to strong radial migration as a result 
of gas drag (Weidenschilling, 1977). For example, in a massive axisymmetric disc, the 
predicted timescale on which metre sized bodies  would be swept into the star from a 
radius of $5$ AU is $\sim$ a few $10^3$ years (cf. Rice et al., 2004) and there is therefore 
the concern that the disc  could be severely depleted in solid material during the early 
self-gravitating  phase of disc evolution. This would be at odds with observational 
evidence from debris discs that at least some objects retain several tens of earth 
masses of rocky/icy debris after the disappearance of the disc gas, and would also severely 
impact  the planet formation potential of the disc.  
 
Rice et al. (2004) however  suggested a way that this outcome could be 
circumvented, by distinguishing the rapid radial migration that
occurs in a massive {\it axisymmetric} disc from the behaviour expected 
when one takes into account the strong {\it spiral} features in massive, 
self-gravitating discs. Their simulations demonstrated that in this case the 
effect of gas drag is to strongly concentrate objects of around metre size in the pressure
maxima associated with spiral shocks. In the simulations, this concentration
resulted in a more than hundredfold increase of the local solid density,
bringing the simulations into the regime, for which they were not designed,
where the disc is locally dominated by the mass in solids rather 
than in gas. Rice et al. (2006) however estimate that the conditions in the
arms are not only conducive to runaway grain growth by collisions
but can also lead to the creation of large (km scale) planetesimals through
the action of self-gravity in the solid phase (see also Goldreich \& Ward 1973, 
Youdin \& Shu 2002). If this is indeed the case, then it effectively `rescues' the 
solids from rapid radial migration, since gas drag is unimportant for such large 
objects. This scenario then raises the possibility of storing some fraction of
the solids in self-gravitating discs  (which initially total several hundred earth
masses of solids) in the `safe' form of planetesimals. In principle, these planetesimals 
could then be available for involvement in  future planet building, as well as providing 
a reservoir for dust production in future debris discs.

Such a picture is only workable, however, if a number of conditions are met,
regarding both the viability of planetesimal formation in self-gravitating
discs, and the subsequent evolution of the planetesimal swarm. In this
Letter we concentrate on the former issue: although Rice et al. (2004)
demonstrated the viability of dust concentration in spiral arms in their
simulations (which employed a simple scale free `toy' cooling model for 
the disc gas), we here explore what are the conditions required for this mechanism to work
in discs with realistic cooling. Our conclusions will be used to provide initial
conditions for future calculations of planetesimal evolution in self-gravitating discs. 

In Section 2 we explain how the viability of solid growth in spiral
features depends on the   dust concentration timescale as compared with the 
(roughly dynamical timescale) lifetime of  individual spiral features. The dust concentration
timescale is linked to the amplitude of spiral features in the gas, which itself depends on the 
cooling timescale of the gas. In section 3 we then use analytic models of the structure of 
self-gravitating discs subject to realistic cooling in order to demonstrate that the concentration 
of dust in spiral features is feasible only in the outer regions of proto-planetary discs (beyond
a few tens of AU).  In Section 4 we discuss the implications of this result for planet formation 
models and for the retention of dust grains in protostellar discs.
We emphasise that other mechanisms may also be efficient in the rapid production
of planetesimals (see e.g Ciesla 2009, Brauer et al 2008, Kretke \& Lin 2007)
and that our conclusions here apply only to the viability of planetesimal
assembly in self-gravitating discs. 

\section{The link between dust concentration efficiency and cooling timescale }

 The success of numerical simulations in achieving the desired concentration
of dust in spiral features is at first sight surprising, given the fact that in self-gravitating 
gas discs it is well established that individual spiral features are transient 
(though regenerative) features (e.g., Lodato \& Rice, 2004, Britsch et al., 2007).
The characteristic lifetime for spiral features is of of order the dynamical
timescale ($\Omega^{-1}$) and so the fact that spiral arms are able
to concentrate dust successfully suggests that the concentration timescale
must also be $< \Omega^{-1}$. Rice et al. (2004) present simple arguments that this 
should indeed be the case;  these arguments however rely on an assumption that
the spiral features involve pressure variations of order unity, which is
indeed the case in these particular  simulations. Nevertheless, our
recent simulations demonstrate that the amplitude of spiral features is itself
a function of disc cooling timescale (Cossins et al., 2009); the observed
dependence (whereby the fractional amplitude of spiral features scales
as the inverse square root of the ratio of cooling timescale to dynamical
timescale) may be simply understood in terms of the properties of weak
adiabatic shocks. This means that in practice {\it the concentration of dust in
spiral features will be restricted to regions of the disc where the cooling
timescale is appropriately short.}

\subsection{The link between dust concentration efficiency and spiral arm amplitude}

  The concentration of dust in spiral arms is effected by the same mechanism
that is involved whenever dust concentrates in pressure maxima in gas discs
(e.g. Haghighipour \& Boss 2003, Godon \& Livio 2000, Klahr \& Bodenheiner 2003):
this same mechanism  has been invoked in cases  where a variety of physical 
processes (e.g. vortices or edges induced by planets/binary companions) 
are responsible for the creation of the local pressure maximum. In each case,
dust concentration relies on the fact that the gas is partially pressure
supported and thus its local orbital frequency is respectively sub (super)-
Keplerian outside (inside) of the pressure maximum. Dust particles
behave ballistically to first order and thus orbit respectively faster (slower)
than the gas outside (inside) the pressure maximum. If one now introduces
drag forces between the dust and the dominant gas component, the dust is then
decelerated (accelerated) by the gas outside (inside) the pressure
maximum and therefore moves inwards (outwards). Thus dust accumulates
preferentially at the location of pressure maxima.

The magnitude of the radial velocity induced by drag on the solid component depends 
on the ratio between the stopping time $t_{\rm s}$ and the dynamical time $\Omega^{-1}$ 
(see, for example, Takeuchi \& Lin, 2002).  In the limit of weak drag forces (i.e where the 
stopping timescale $t_{\rm s}$ is much longer than the local orbital time $\Omega^{-1}$) 
then the rate of radial migration simply increases with increasing drag strength (i.e.
it scales as $1/\Omega t_{\rm s}$). If however the drag is very strong ($\Omega t_{\rm s} <<1$) 
then the rate of radial migration is now set by the terminal velocity of the grain's radial flow, 
which now scales as $\Omega t_{\rm s}$ (i.e. decreases with increasing drag strength). We thus have
\begin{equation}
v_{\rm rad} \sim \frac{\Delta v}{\Omega t_{\rm s} + 1/\Omega t_{\rm s}}.
\end{equation}
The maximum radial velocity is acquired at intermediate drag strength when
$\Omega t_{\rm s} \sim 1$, which corresponds roughly to particles of metre size
in the case of the massive (self-gravitating) discs considered here
(Rice et al 2004).  
Since in this case the grain is brought into co-rotation with the 
gas over each orbital period, then the radial velocity acquired is roughly  the difference in 
orbital velocity  ($\Delta v$) between a ballistic grain and the local gas velocity.  

This velocity difference $\Delta v$ can be readily assessed by consideration
of radial force balance for the gas, where the difference between centripetal
acceleration and gravitational acceleration is provided by the  acceleration
due to the local pressure gradient. Thus
\begin{equation}
v_\phi^2 = v_K^2 +{R\over\rho}{\partial P\over \partial R},
\end{equation}
where $v_K$ is the Keplerian speed at which the dust orbits
and $v_\phi$ is the orbital speed of the gas, such that it experiences
a net centrifugal force that matches the combination of gravitational
and radial pressure forces. In the case that we have a surface density enhancement
$\sim \Delta \Sigma$ over a radial length scale $\lambda$ we can approximate
the second term on the right hand side as   $\sim c_s^2 (R/\lambda)(\Delta\Sigma/\Sigma)$ 
(where $c_s$ is the sound speed). Thus
\begin{equation}
v_{\phi}^2 = v_K^2 \left [ 1+\left(\frac{c_s}{v_K}\right)^2 \frac{\Delta \Sigma}{\Sigma}\frac{R}{\lambda}\right ]
\end{equation}
and since the second term on the right hand side is $<< 1$, we may write
\begin{equation}
\Delta v = v_\phi - v_K \sim v_K \left(\frac{H}{R}\right)^2 \left(\frac{R}{\lambda}\right) \left(\frac{\Delta \Sigma}{\Sigma}\right)
\end{equation}
where we have used the thin disc relation $H/R \sim c_s/v_{\rm K}$. Therefore (since the maximum 
radial inflow rate is $\Delta v$) we find  that the minimum time $t_{\rm min}$ required to 
concentrate solid material in a pressure maximum is $\sim \lambda/\Delta v$ and thus:
\begin{equation}
t_{\rm min} = \lambda/\Delta v = \Omega^{-1} (\lambda/H)^2 (\Delta \Sigma/\Sigma)^{-1}
\end{equation}
 
Since in a gravitationally unstable disc the scale length for density inhomogeneities is of 
order the most unstable wavelength, which, in marginally unstable discs is of order $H$ 
(Toomre 1964, Lodato 2007), we can set $\lambda \sim H$ and hence
\begin{equation}
t_{\rm min} = \Omega^{-1} \left(\frac{\Delta \Sigma}{\Sigma}\right)^{-1}
\end{equation}

\subsection{The link between spiral arm amplitude and cooling timescale}

Cossins et al. (2009) established, through analysis of a suite of simulations
with various imposed ratios of cooling timescale to dynamical timescale,
that the fractional amplitude of spiral features scales as the inverse
square root of the cooling time:
\begin{equation}
\frac{\Delta \Sigma}{\Sigma} = {{1}\over{\sqrt{\Omega t_{\rm cool}}}}
\end{equation}

This result can be simply understood, given that Cossins et al. (2009) also 
showed that the shocks in these discs are only marginally supersonic 
(i.e. with Mach number $=1 + \epsilon$, where $\epsilon \ll 1$).
The cooling timescales in gravitationally unstable discs are sufficiently
long that cooling can be neglected {\it at the shock front} and thus we
can invoke the result for the vertically averaged density contrast in
weak adiabatic shocks:
\begin{equation}
\Delta \Sigma /\Sigma \propto  \epsilon
\end{equation}
Nevertheless
cooling becomes important downstream of the shocks: in a state of thermal
equilibrium we have the situation where the conversion of mechanical
energy into heat that is achieved at the shock front is balanced by cooling
downstream of the shock. The quantity of energy dissipated at the shock
is proportional to the entropy jump at the shock which, in the case
of weak adiabatic shocks, scales as $\epsilon^2$. We thus expect, in thermal
equilibrium that
\begin{equation}
\epsilon \propto 1/ \sqrt{t_{\rm cool}}
\end{equation}
and thus, combining equations (8) and (9)
\begin{equation}
\Delta \Sigma/\Sigma \propto 1/ \sqrt{t_{\rm cool}}
\end{equation}
in agreement with the empirical result equation (7).

\section{The efficiency of dust aggregation in discs with realistic cooling}

The results of Section 2 (equation (6)) imply that the minimum timescale for 
dust aggregation (normalised to the local orbital timescale) is of order the fractional
density enhancement in the shock and this scales with the inverse square
root of the ratio of the cooling timescale to dynamical timescale, as given
by equation (7). We thus have
\begin{equation}
\Omega t_{\rm min} \simeq  \sqrt{\Omega t_{\rm cool}} 
\end{equation}

On the other hand, the lifetime of individual spiral features in 
self-gravitating discs is of order $\Omega^{-1}$. We thus see that
the extent to which dust can accumulate in a given spiral arm  before this
feature dissolves is related to the fractional density enhancement in
the shock and hence on the local cooling timescale. 

  Figure 1 depicts contours of constant fractional density enhancement
(and hence $\Omega t_{\rm cool}$) in the plane of radius versus steady state
accretion rate for a disc around a solar mass object. These contours are
constructed on the assumption that the transport properties of a
marginally self-gravitating disc can be described as a pseudo-viscous process
(see e.g. discussions in Cossins et al., 2009, Clarke 2009): specifically
this allows one to relate the rate of energy dissipation to the local
angular momentum transport (and hence radial mass flux) in the disc
(see also Rafikov 2009). In
a state of thermal equilibrium, this energy dissipated is balanced by cooling
(on a timescale $t_{\rm cool}$) so that (if one knows $t_{\rm cool}$ as a function
of local conditions) one can relate $t_{\rm cool}$ to a local (pseudo-)viscosity.
In Figure 1, we have simply replaced this local pseudo-viscosity by the
corresponding accretion rate in the case that the disc was in a steady state:
although this parameterisation depends on the steady state assumption, the
local solutions (i.e. the relationship between local density, temperature
and pseudo-viscosity in a given cooling regime) are not restricted to the
steady state case. Our parameterisation however allows one to interpret 
solutions in terms of a quantity (i.e. accretion rate) that can be
estimated in young stars. 

  The solutions on which Figure 1 is based assume that the disc cools
radiatively: for the parameters of interest, the disc is optically thick
and we hence model cooling via the radiative diffusion approximation,
with the (Rosseland mean) opacity given by the piece-wise power law fit
as a function of density and temperature given by Bell \& Lin (1994).  Figure 1 
shows the division of the $(\dot M, R)$ plane according to the dominant opacity 
source (marked by dashed lines) and delineates the non-fragmenting marginally 
unstable regime by  the two bold lines. To the right of this region, the ratio of 
cooling timescale to dynamical timescale is sufficiently short that the disc 
fragments (see e.g. Gammie 2001, Rice et al., 2005), rather than remaining in a 
self-regulated state of marginal gravitational  stability: this line corresponds to the 
case where the fractional amplitude is about unity (Cossins et al., 2009). To the left,  
the dominant angular momentum transfer is provided by the magneto-rotational 
instability (MRI) and the disc is non-self gravitating. The dotted line indicates the 
contour where the fractional amplitude $\Delta\Sigma/\Sigma$ is 10 percent.
For details concerning the construction of Figure 1, together with analytic 
expressions for the equilibrium  solutions in various regimes and a discussion of 
our assumptions about the regions of the disc where the MRI is effective, 
see Clarke (2009). 

\begin{figure}
\vspace{2pt}
\epsfig{file=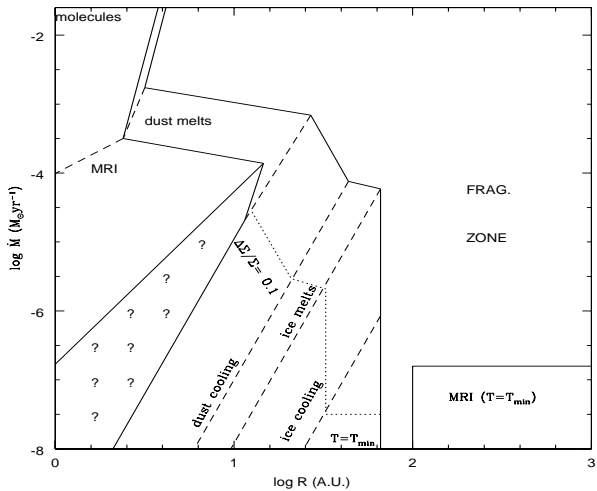,width=8.5cm,height=7.cm}
\caption{Regimes in the plane of steady state accretion rate versus
radius in the case of a disc surrounding an object of mass $1 M_\odot$,
showing the dotted contour where the fractional surface density amplitude is
$10 \%$. The bold lines that parallel this dotted contour to the  upper right
represent the condition of unit fractional amplitude and also marks
the boundary between fragmenting and non-fragmenting conditions in
the disc gas. The region denoted with question marks is a region where the disc is expected 
to be too cold in order for the MRI to be active, but where a self-gravitating disc would be too hot for 
self-gravity to dominate angular momentum transport (full details can be found in Clarke, 2009).}
\end{figure}

We immediately see that the regions of the disc where large amplitude
spiral features are to be expected are located at rather large radius
(i.e. many tens of AU). Note the general argument given above that effective 
dust concentration implies that $\Delta \Sigma /\Sigma$ must not be 
{\it much} less than unity. The exact minimum value for $\Delta \Sigma /\Sigma$ required to 
induce substantial dust concentration cannot be obtained from the simple order of magnitude  
estimates provided above and needs to be obtained through detailed numerical
simulations. Rice et al. (2004) obtain effective dust concentration in simulations where 
$\Omega t_{\rm cool} = 7.5$, which corresponds to $\Delta \Sigma /\Sigma\approx 0.1$, thus placing 
an upper limit of around 10\% to such minimum value. Unfortunately, it is not currently possible 
to perform self-consistent simulations of self-gravitating disc with $\Omega t_{\rm cool}$ 
greater than around $10$, because at this point the transport of angular momentum and 
dissipation of energy becomes dominated by numerical viscosity rather than self-gravity. 
Thus, in Figure 1 we show the line where the fractional amplitude is $10 \%$ as a guide to the
region of the disc beyond which dust aggregation is likely to be
effective. We note that the locus of constant fractional amplitude is
independent of accretion rate for accretion rates over a range of values that
are appropriate to young low mass stars (i.e. from   $10^{-6} M_\odot$ 
yr$^{-1}$ to   $< 10^{-7} M_\odot$ yr$^{-1}$):  this is because the dominant
opacity source is provided by ice grains, for which the opacity scales
as $T^2$ and in this case the cooling timescale for a disc that
is marginally gravitationally unstable turns out to be a function of radius
only, independent of temperature (or accretion rate).  

\section{Discussion}

  We have shown that the concentration of dust in spiral features is
likely to be viable only in the outer regions of proto-planetary discs
(i.e. beyond a few tens of AU), even though the disc may be self-gravitating
at smaller radii.  The inner extent of the zone of the disc where 
this mechanism is effective lies at a radius that is $\sim 3-10$ times
smaller than the innermost radius where gas phase fragmentation of the disc
is possible.  We note that the recently imaged planets in nearby young stars
(Marois et al 2008, Kalas et al 2008) lie at orbital radii that would be
consistent with the planetesimal creation mechanism discussed here.
 
We base our  conclusion about viable regions of the disc
on the fact that dust will only
concentrate in spiral features if its  concentration timescale is less
than or of order the lifetime of individual features, which itself is
of order the local orbital timescale. A short concentration timescale
requires that the fractional amplitude of density enhancements in spiral arms
is not much less than unity and this in turn implies that the local ratio
of cooling timescale to dynamical timescale is sufficiently short. 
Figure 1 shows that the region of the disc that satisfies this requirement
is at rather large radius and that  the extent of this region is not
a strong function of accretion rate in the disc. We emphasise that Rice
et al. (2004) obtained successful dust concentration in spiral features because
their toy cooling model was set up with an appropriately short cooling
timescale. 

  As set out in the Introduction, the fact that solid material in the outer
regions of self-gravitating discs may be efficiently converted into 
planetesimals may be critical for the retention of solids in the disc, 
since such planetesimals are immune to the migration effected  by gas 
drag in the case of smaller grains. Pilot simulations of the dynamical 
behaviour of planetesimals in self-gravitating discs (Britsch et al., 2008) 
imply that the planetesimal orbits are
driven to moderate eccentricities (of order $0.1$) by interaction with
fluctuating spiral features in the gas disc: the relatively large
resultant velocity dispersion (of order a km/s), means that the runaway
growth of this population in the direction of planet building is unlikely
while the disc is still self-gravitating, but these velocities do suggest
that collisions, when they occur, will be destructive and that small
grains may be re-populated in the disc in this way (note that, since small
dust is closely coupled to the gas, then the regenerated dust could then in 
principle re-enter the cycle of grain growth and  - as long as the disc
remains self-gravitating - accumulation in spiral shocks).
Observationally,
mm emission is detected in discs over many millions of  years (i.e. long
after the disc ceases to be self-gravitating), and shows  no
appreciable decline with time over this period
(Andrews \& Williams, 2005). Given the short radial migration timescales
for such grains (less than a million years), this fact requires that they
are replenished from some reservoir (see Takeuchi et al., 2005, Wyatt et al.,
2007). The planetesimals assembled in the outer disc during the previous
self-gravitating phase may provide a suitable stock pile of material
from which small grains could be collisionally replenished. Before exploring
such ideas in more detail (see e.g. Garaud, 2007) it is however first necessary
to demonstrate that these planetesimals can be retained in
the disc when subject to the dynamical evolution
that results from  interaction with the self-gravitating disc.

 This issue of dynamical evolution also impacts on the other 
possible observational consequence of early planetesimal creation, i.e.
the location of dust in debris disc systems.  We have shown that we expect
the creation of planetesimals in self-gravitating discs to be limited to
large radii (i.e. a scale of tens of AU or more), which is a region similar
in size to the planetesimal belts inferred in debris disc systems (Wyatt, 2008).
(Note that, as mentioned in the Introduction, there are other candidate
mechanisms for planetesimals assembly which may be operative in different
regions of the disc: for example, some scenarios place the location of
planetesimal assembly close to the ice-line, i.e. at a radius of $\sim 3$ A.U.
in the case of solar type stars: see Kretke \& Lin 2007, Brauer et al 2008). The
mass in planetesimals that is deduced in the  case of observed debris
discs around sun-like stars is several tens of
earth masses and a similar figure is deduced from the requirement that
Pluto and similar bodies can form on the same timescale as Neptune (Stern
and Colwell 1997, Kenyon and Luu 1999),
which necessarily implies formation in the early, gas rich phase
of disc evolution. {\footnote {A similar mass is also inferred in the case of the
{\it primordial} Kuiper belt, from which it is then argued that the much smaller
mass of the current belt is a consequence of depletion in a  dynamical
instability  occurring up to a Gyr later: Gomes et al 2005}}. Such masses
represent  a relatively small fraction of the total solid mass
in the disc during the self-gravitating phase (which is of order hundreds
of earth masses), implying that the planetesimal belts
in debris disc systems may be a remnant of an 
initially larger population of
planetesimals that were assembled in the self-gravitating phase. Again,
further dynamical simulations are required in order to ascertain whether
the ultimate distribution of planetesimals would retain a memory of their
initial creation sites.  

 If such a memory
were  retained, then this would imply an inner hole in the planetesimal
distribution at the time that the disc gas was dispersed, i.e. at the
beginning of the debris disc phase. It is hard to establish whether or not 
debris discs initially contain planetesimal belts 
that extend in to  small  radii, since  the short
collisional lifetime at small radii means that the resultant dust    
should not be observable for long (Wyatt 2008). Current surveys
(Carpenter et al 2009)
have failed to detect  hot dust (as evidenced by $8 \mu$m emission)
in optically thin discs around
solar type  stars, but the relatively small number of young stars in the sample
make it impossible to conclude from this that planetesimal belts at
a scale of $\sim 1$ A.U. are necessarily absent in debris discs at birth.
The statistics of debris discs with cool dust (associated with $24 \mu$m
emission) also admits a variety of interpretations. According to the models
of Kenyon and Bromley (2005), a disc with an inner hole in its planetesimal
distribution should produce a {\it delayed} rise in (cool) dust production,
since one has to wait for the formation of large planetesimals that
can initiate a collisional cascade in smaller bodies. The formation timescale for large ($\sim 2000$ km) planetesimals at $30$ A.U. is relatively long
($\sim 10$ Myr), so that according to this argument, debris should
be absent in 
systems younger than  $\sim 10$Myr if primordial holes in the planetesimal
distribution are common. However, 
the handful of $24\mu$m detections in younger systems
(Carpenter et al 2009) does not  rule out this scenario, 
since 
these few objects could instead be the final  remnants of 
primordial discs, where dust does not trace the location of 
a population of colliding planetesimals.
Larger surveys studying the evolution of  debris in  younger stars 
are needed in order to 
constrain the  architecture of planetesimal belts in solar type stars: 
it is notable that in the case of debris discs around A stars (where
larger samples are available) there is some claimed evidence for a
maximum brightness of debris at an age of $ 10-15$Myr (Currie et al 2008), 
which would be compatible
with a $\sim 30$ A.U. hole in the planetesimal distribution in these objects.

 In conclusion, we remark that it is often assumed that the spatial
location of planetesimal belts is set by the orbital parameters
of giant planets (see e.g. Quillen 2006 and the recent striking
confirmation in the case of the Fomalhaut system of a
suitably located planet in the debris disc: Kalas et al 2008). 
Here we merely point out  
that the location of
planetesimal belts may also 
relate to the
initial mechanism for planetesimal production. We have shown in this paper
that the creation of planetesimals in self-gravitating discs is an outer disc
phenomenon.  The implication of this result for planet formation and
the evolution of debris discs is yet to be explored.    

\section{Acknowledgments}
We thank Mark Wyatt for constructive comments and the anonymous referee
for providing us with a number of useful references.


\bibliographystyle{plainnat}

\label{lastpage}
\end{document}